\nonstopmode
\documentstyle[11pt,paspconf]{article}

\begin{document}

\title{A Retrospective View of Miriad}

\author{R.J. Sault
}
\affil{Australia Telescope National Facility, CSIRO, Epping, Australia}

\author{P. J. Teuben
}
\affil{Astronomy Department, University of Maryland,
	College Park, MD}

\author{M.C.H. Wright
}
\affil{Astronomy Department, University of California,
    Berkeley, CA}

\begin{abstract}
Miriad is a radio interferometry data-reduction package aimed at
taking raw data through to the image analysis stage. The Miriad
project, begun in 1988, is now middle-aged. With
the wisdom of hindsight, we review design decisions and some of Miriad's
characteristics.
\end{abstract}

\keywords{software, interferometry, visualization}

\section{A Brief History}
Miriad aims at being a ``full service'' radio interferometry data-reduction
package, taking raw telescope data through to image analysis
and publication-quality image displays. The Miriad
project was initiated by the Berkeley Illinois Maryland Association
(BIMA). This followed much internal debate, and a
meeting in February 1988 where a number of external experts (from the AIPS,
GIPSY and IRAF camps) were invited to express their opinions on
BIMA's offline software options. BIMA's decision was to develop
two streams. The first stream, the Miriad project, was to build on the
experience with the Illinois \verb+Werong+ and Berkeley \verb+RALINT+ packages, and
to develop a package up to, but excluding, the image analysis stage. The
second, image-analysis, stream eventually died, and Miriad
was extended to cover this area.

Miriad was designed and developed jointly by groups at the three BIMA
sites.  Most of the infrastructure was written in Illinois by one person,
with significant input from Berkeley.
The first astronomically useful applications 
(FITS readers/writers, imaging and deconvolution tasks)
appeared in October 1988. 
A fourth site became involved when two of the Miriad
group moved to the Australia Telescope National Facility in 1990, and began
extending Miriad to meet the needs of that institute's interferometer.

\section{Project Management and Economics}
Having the project spread over three or four sites has its
difficulties, especially when
group members are responsible to individual
institutes and not to the project as a whole. In the early stages,
there were regular
phone conferences, quarterly
face-to-face meetings and a
barrage of email. There was no real project leader, with the project
advancing
by the good will of the group
members (a true anarchy). The pragmatic solution to the
occassional disagreements
between group members (and the interests they serve) was nick-named
``the free market economy'' (a true capitalism). Each
piece of code has an ``owner'' who must approve any change to it.
If an owner cannot be persuaded, anyone is free to submit
alternative code, and the different codes compete in the open market for
users. Generally this occurs only occassionally, but it has led
to multiple, functionally similar, tasks of significantly
varying quality. This
``free market'' approach results in some inefficient use
of resources which perhaps could be avoided with a ``centralised
economy''.

\section{User Interface and Documentation}
User interfaces were a controversial issue in the early stages of the
project -- no interface suits every use. The approach
adopted was to make the ``front-end'' user interface quite separate from
number-crunching tasks.
Apart from running tasks, the main functions of a front-end program
is to provide task
documentation and to help assemble task parameters. These
parameters are passed to the number-crunching tasks simply as command
line arguments.
This trivial interface made it easy to develop front-end
interfaces to suit different tastes; early front-end programs included
a Sunview
windowed environment, a VT100 menu system, and a ``dumb terminal''
interface similar to POPS.

This design means that the front-end task can be completely
bypassed, with users initiating tasks at the host
command line. More experienced users do this often, particularly
with simpler tasks. It is also the best way to develop
batch scripts (we prefer to write those in the
powerful shell that the host 
provides, rather than to try to reinvent it).

Using an idea adapted from PGPLOT,
task documentation is stored as comments as a
preamble in the source code.
A tool extracts this `help file' and stores it in a
directory ready for use by the front-end programs
(whenever a task is recompiled, its help file is also updated).
Help files for subroutines are treated in a similar fashion.

Although help files are good for specific information, additional
documentation giving a greater overview is needed.  Users and
programmers guides appeared early in Miriad's life. Later a
guide specifically aimed at Australia Telescope users was also developed.
Although this is continually updated, re-assignments within BIMA
have meant that their guides are now out-of-date. 

With the popularity of the WWW, a html version of Miriad's documentation
was developed (e.g. see \verb|http://bima.astro.umd.edu/bima/miriad| or
\verb|http://www.atnf.csiro.au/ATNF/miriad|). This hypertext version is
generated automatically from the source of the users guide and the help files.

\section{File Format Issues}
Compatibility with FITS is clearly an important goal.
Although we considered the possibility of
making FITS the `native' file format of Miriad (as it was in \verb+Werong+),
we concluded that pure FITS was too inflexible for
this purpose. However, we adopted an image data-set structure which
shadowed FITS reasonably closely (thus making a translation between the
two straightforward). This was not
possible with visibility data-sets -- the general characteristics of the uv FITS
format were seen as too restrictive for our needs. The visibility
format adopted is based on the \verb+RALINT+ design
of Wilson Hoffman. It has proven to be significantly more flexible than
the FITS style of handling visibility data, and has the added advantage
of a much cleaner interface for the programmer.
We have found it easy to use this format with a
number of forms of data not considered in the original design.
Unfortunately it has the penalty that the current implementation can
be slow to read.

Our initial concept of Miriad was of a package working in a 
shared disk environment of VAXes, Suns and Crays -- machine-independence
of the data was seen as very important. Miriad
data are stored in a canonical format on disk, with
the i/o system performs needed conversions during the i/o
process (this conversion is invisible to the applications programmer).
Although this has been very successful,  because of declining interest
in VAXes and Crays, machine independence has not been
as critical as we had anticipated.

\section{Language and Portability Issues}
Miriad was designed to be portable, with VMS and several UNIX variants
being the initial target systems.
Since then, Miriad has been ported, with minimal effort, to many
UNIX-based systems. 
To aid portability, and to reduce our own workload, we have gladly
used public-domain software where appropriate. Miriad's plotting
tasks are based on PGPLOT, whereas some of the numerical code is
based on LINPACK.

Although all the i/o system and ancillary tools are written
in C, most of Miriad is written in FORTRAN. We used
FORTRAN because we felt that 
most astronomers would be happiest with it (and we wanted
to attract programmers), and because vector machines
were important to us (the best vectorising compilers continue to
be FORTRAN ones).

As the i/o system is written in C, at some level FORTRAN has to call
C routines. This language barrier is invariably system
dependent (there are six schemes used in the systems that Miriad
has been ported to). We developed a tool which
takes a system-independent interface description, and
which produces a thin layer of system-dependent code that mates the FORTRAN
and C parts of Miriad.
Although we have always had some misgivings about a mixed-language system, this
approach has worked reasonably well.

\section{Visualisation}
Visualisation and image display is one of Miriad's failings. In the
initial stages, this area was split off to a sub-group to develop. Their
plans were comprehensive and their development lagged behind the rest of
the project. As a stop-gap, a simple interim set of routines was
adopted. Eventually, the comprehensive plan failed and that sub-group disbanded.
Somewhat later, a second attempt was made. Although this did progress further, it
also effectively failed, and this second sub-group also disbanded.

Meantime, the ``interim'' routines were slowly upgraded, and made to
work under X-Windows, but they are still basic. These
shortcomings have been somewhat alleviated by
the development of a set of image display tasks using
PGPLOT. Though a good plotting package, PGPLOT is not intended to be an
image display package, and so its model of a display device is
limiting.

\section{Miriad's On-Line Component}
One novel aspect of Miriad is that
it has been integrated into the on-line system of the Hat Creek
interferometer. 
The on-line
system generally uses the Miriad user interface and documentation system.
Thus, to some
extent, the user interface remains the same from driving the
telescope to producing publication quality output.
The on-line system also writes the raw data directly in the Miriad uv
data format (this requirement was one of the reasons uv FITS was
unsatisfactory).

\section{A Niche Package}
Miriad is continuing to be developed.
It has proven to be a good and flexible environment for writing
specialised applications, as well as for developing
new algorithms of greater applicability. 
A natural question to ask is ``why reinvent the
wheel''. Certainly at the original planning meeting there was a strong
voice (both from NRAO and some BIMA representatives) for using AIPS to
solve BIMA's reduction problems. AIPS, however, did not
satisfy a number of the criteria that BIMA felt were essential.
We had the choice between a major development in AIPS, or a
major development with a fresh system. A fresh start, a system
more flexible than AIPS, and a more programmer-friendly environment were
probably the deciding factors (politics may have also played a part).

In hindsight, it was the correct decision. We had a useful system
comparatively quickly, and have been able to extend it with new
algorithms and techniques at a good rate. The overhead of programming in
AIPS would have, at best, slowed our software development. At worst, it would
have completely dissuaded us from implementing many new applications.

Faced with AIPS++, what is the future of Miriad? Part of the success of
Miriad is that it is not a huge package -- it has been able to
adapt and concentrate on specialised areas, and in a timely fashion.
Miriad does not try to
address the data-reduction needs of the entire radio-astronomy
community, so
the overheads are much less than those
for AIPS++. We do not believe that mega-packages will, or should,
swamp the small and mid-size packages; there will always be a place for
these. At the same time
Miriad is now showing some grey hairs -- perhaps Miriad++ is needed.

\acknowledgments

We thank the many people who have helped make the package successful, 
in particular (in historical order)
Wilson Hoffman, Brian Sutin, Lee Mundy, Neil Killeen, Jim Morgan, Bart Wakker
and Mark Stupar.

\end{document}